\documentclass[twoside]{article}
\usepackage{fleqn,espcrc2}

\usepackage{graphicx}
\usepackage[figuresright]{rotating}

\newcommand{\be}{\begin{equation}}
\newcommand{\ee}{\end{equation}}
\newcommand{\ba}{\begin{eqnarray}}
\newcommand{\ea}{\end{eqnarray}}

\newcommand{\bx}{{\bf x}}

\newcommand{\matrixele}[3]{\ensuremath{\langle #1 \mid #2 \mid #3\;\rangle}}
\newcommand{\AmS}{{\protect\the\textfont2
  A\kern-.1667em\lower.5ex\hbox{M}\kern-.125emS}}

\title{Scalar and Tensor Glueballs on Asymmetric Coarse Lattices}

\author{C. Liu\address{Department of Physics, 
        Peking University, 
        Beijing 100871, P.~R.~China}%
        \thanks{Work supported by the Chinese Natural Science Foundation
                under Grant No. 19705001, the Climb-up Fund from Ministry
                of Science and Technology and the Startup fund from Peking
                University.}
       }
       
\begin{document}

\begin{abstract}
Scalar and tensor glueball spectrum 
is studied using an improved gluonic action on
asymmetric lattices in the pure $SU(3)$ gauge theory. The smallest spatial
lattice spacing is about $0.08fm$ which makes the extrapolation to the
continuum limit more reliable.
In particular, attention is paid to the scalar glueball
mass which is known to have problems in the extrapolation.
Converting our lattice results to physical units using the
scale set by the static quark potential,
we obtain the following results for the glueball masses:
$M_G(0^{++})=1730(90)MeV$ for
the scalar glueball mass and $M_G(2^{++})=2400(95)MeV$ for
the tensor glueball.
\vspace{1pc}
\end{abstract}

\maketitle

\section{INTRODUCTION}
\label{sec:intro}

It is believed that QCD is the theory which describes
strong interactions among quarks and gluons. A direct consequence of
this is the existence of excitations of pure gluonic degrees of freedom,
i.e. glueballs.  However, due to their non-perturbative nature,
the spectrum of glueballs can only be investigated reliably with
non-purterbative methods like lattice QCD %
\cite{bali93,michael89,berg83a,berg83b,%
schierholz82,schierholz83}. Recently, it has become clear
that such a calculation can be performed on a relatively coarse lattice
using an improved gluonic action on asymmetric lattices %
\cite{lepage95:pc,colin97,colin99}. In this paper,
we present our results on lowest scalar and tensor glueball 
spectrum calculation.
The spatial lattice spacing in our simulations ranges from $0.08fm$ to
$0.25fm$ which enables us to extrapolate more reliably to the
continuum limit.
The improved gluonic action we used is the tadpole improved gluonic
action on asymmetric lattices as described in \cite{colin97,colin99}.
It is given by:
\ba
S=&-&\beta\sum_{i>j}
\left[{5\over 9}{TrP_{ij} \over \xi u^4_s}
-{TrR_{ij} \over 36 \xi u^6_s}
-{TrR_{ji} \over 36 \xi u^6_s} \right] \nonumber \\
&-&\beta\sum_{i} \left[
{4\over 9}{\xi TrP_{0i} \over  u^2_s}
-{1\over 36}{\xi TrR_{i0} \over u^4_s} \right] \;\;.
\ea
In the above expression, $\beta$ is related to the bare gauge coupling,
$\xi=a_s/a_t$ is the (bare) aspect ratio of the asymmetric lattice with
$a_s$ and $a_t$ being the lattice spacing in spatial
and temporal direction respectively. The parameter $u_s$ is the
tadpole improvement parameter to be determined self-consistently from
the spatial plaquettes in the simulation. $P_{ij}$ and $P_{0i}$ are the
spatial and temporal plaquette variables.
$R_{ij}$ designates the $2\times 1$
Wilson loop ($2$ in direction $i$ and $1$ in direction $j$).
Using spatially coarse and temporally fine lattices  helps
to enhance signals in the glueball correlation
functions. Therefore, the bare aspect ratio
is taken to be some value larger than one. In our simulation, we
have used $\xi=3$ for our glueball calculation.
It turns out that, using the non-perturbatively
determined tadpole improvement \cite{lepage93:tadpole}
 parameter $u_s$, the renormalization
 effects of the aspect ratio is small \cite{colin97,colin99},
 typically of the order of a few percent for practical
 values of $\beta$ in the simulation, which is 
 ignored in this paper.


\section{MONTE CARLO SIMULATIONS}

We have utilized a Hybrid Monte Carlo algorithm to update gauge
field configurations. Several lattice sizes have been
simulated and the detailed information can be found in Table.1.
\begin{table}[htb]
\caption{Simulation parameters and the corresponding
lattice spacing in physical units obtained from Wilson loop
measurements. Parameters used in the smearing process for
Wilson loop measurements are also listed.}
\vspace{3mm}
\begin{center}
\begin{tabular}{@{}lllll}
\hline
      Lattices & $\beta$ & $\lambda_W$ & $n_W$ & $r_0/a_s$  \\
      \hline
      $ 8^3\times 24$ & $2.4$ & $0.20\sim 0.40$ & $2\sim 4$ & $1.98(2)$ \\
      $ 8^3\times 24$ & $2.6$ & $0.20\sim 0.40$ & $2\sim 4$ & $2.48(2)$ \\
      $ 8^3\times 24$ & $3.0$ & $0.20\sim 0.40$ & $4\sim 6$ & $4.11(4)$ \\
      $ 8^3\times 24$ & $3.2$ & $0.25\sim 0.50$ & $4\sim 6$ & $5.89(8)$ \\
      $10^3\times 30$ & $3.2$ & $0.25\sim 0.50$ & $4\sim 6$ & $5.89(8)$ \\
      \hline
\end{tabular}
\end{center}
\end{table}
For each lattice with fixed bare parameters,
order of a few thousand configurations have been accumulated.
Each gauge field configuration
is separated from the previous one by several Hybrid Monte
Carlo trajectories, typically $5\sim 10$, to make sure that
they are sufficiently de-correlated. Further binning of
the data has been performed and no noticeable
remaining autocorrelation has been observed.

\subsection{Setting the scale}
\label{sec:wilsonloop}
In our simulation, the scale is set by measuring
Wilson loops from which the static quark anti-quark potential
$V(R)$ is obtained.
Using the static potential between
quarks, we are able to determine the lattice spacing in
physical units  by measuring $r_0$, a pure gluonic scale determined from
the static potential \cite{sommer94:r0}.
The definition of the scale $r_0$ is given
by: $R^2dV(R)/dR|_{R=r_0}=1.65$.
In physical units, $r_0$ is roughly $0.5fm$
which is determined by comparison with potential models.
For a recent determination of $r_0$, please consult Ref.~\cite{sommer98:r0}.

In order to measure the Wilson loops accurately, it is
the standard procedure to perform single link smearing
\cite{michael89,colin97} on the spatial links 
and  to utilize thermally averaged temporal links of the configurations.
The smearing scheme can be performed iteratively on
the spatial links of gauge fields for as many as
$n_W$ times with a smearing parameter $\lambda_W$.
The smearing parameters $\lambda_W$ and $n_W$ used in this process are
also listed in Table. 1. 
Wilson loops are then
constructed using these smeared spatial links and thermally
averaged temporal links. For a Wilson loop
of size $R\times T$, it is fitted against:
\be
W(R,T) \stackrel{T \rightarrow \infty}{\sim} Z(R) e^{-V(R)T}
\;\;.
\ee
The static quark potential $V(R)$ is obtained from
the effective mass plateau in temporal direction.
Wilson loops measured  along different
lattice axis and at different lattice spacings are seen 
to lie on a universal
line which is an indication that the improved action restores
the rotational symmetry quite well.

The static quark potential is fitted according to
a Coulomb term plus a linear confining potential which is known
to work well at these lattice spacings \cite{colin99}. The potential is
parameterized as:
\be
V(R) = V_0 +e_c/R +\sigma R\;\;.
\ee
From this and the definition of $r_0$, it follows that:
\be
r_0/a_s=\sqrt{{1.65+e_c \over \sigma a^2_s}}
\;\;.
\ee
To convert the measured result to $r_0/a_s$, we have
also used the value of $\xi$ taken as the bare value.
The results of the spatial lattice spacing in
physical units are also included in Table. 1.
The errors for the ratio $r_0/a_s$ are obtained by blocking
the whole data set into
smaller blocks and extracting the error from different blocks.

\subsection{Glueball mass measurement}
\label{sec:correlation}
To obtain glueball mass values, it is
necessary to constructed glueball
operators in various symmetry sectors of interests.
Scalar glueball is
in the $A_1$ representation of the cubic group; tensor glueball
is in representation $E+T_2$, which forms a $5$-dimensional
representation; vector glueballs are in the representation $T_1$
\cite{johnson,berg83a}.

Glueball correlation functions are noisy and difficult to measure.
In order to enhance the signal of glueball correlation functions,
smearing and fuzzying have to be performed on spatial links of the
gauge fields \cite{michael89,colin97,colin99}.
These techniques greatly enhance the overlap of
the glueball states and thus provide possibility of measuring
the mass values.
The smearing and fuzzying process involves typically $6-12$ single link
smearings plus a double link fuzzying.
\begin{figure}[htb]
\begin{center}
\includegraphics[width=6.0cm]{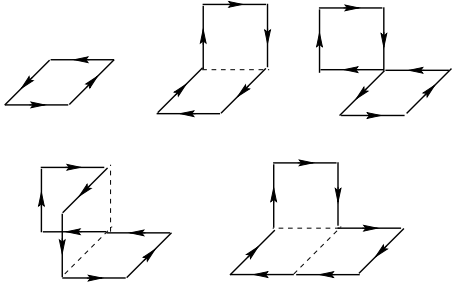}
\caption{The Wilson loop shapes used in constructing the glueball operators.}
\end{center}
\end{figure}
In our simulations, after performing single link
smearing and double link fuzzying on spatial links,
we first construct raw operators at a given time slice $t$:
 $\{ {\mathcal R}^{(i)}_t=\sum_{\bx}{\mathcal W}^{(i)}_{\bx,t}\}$,
 where ${\mathcal W}^{(i)}_{t,\bx}$ are closed Wilson loops
 originates from a given
 lattice point $x=(t,\bx)$. The loop-shapes studied in this
 calculation are shown in Fig.1. Then, elements from the cubic group
 is applied to these raw operators and the resulting
 set of loops now forms a basis for a representation of the cubic
 group. Suitable linear combinations of these operators are
 constructed to form a basis for a particular irreducible
 representation of interest \cite{berg83a}.
 We denote these operators as $\{ {\mathcal O}^{(R)}_\alpha(t) \}$ where
 $R$ labels a specific irreducible representation and $\alpha$ labels
 different operators at a given time slice $t$.

In order to maximize the overlap with one glueball state, we construct
a glueball operator
${\mathcal G}^{(R)}(t)=%
\sum_{\alpha}v^{(R)}_\alpha\bar{{\mathcal O}}^{(R)}_\alpha(t)$,
where $\bar{{\mathcal O}}^{(R)}_\alpha(t)=%
{\mathcal O}^{(R)}_\alpha(t)-\matrixele{0}{{\mathcal O}^{(R)}_\alpha(t)}{0}$.
The coefficients $v^{(R)}_\alpha$ are determined from a variational
calculation. To do this, we construct the correlation matrix:
\be
{\mathbf C}_{\alpha\beta}(t)=
\sum_{\tau}\matrixele{0}
{\bar{{\mathcal O}}^{(R)}_\alpha(t+\tau)\bar{{\mathcal O}}^{(R)}_\alpha(\tau)}
{0} \;\;.
\ee
The coefficients $v^{(R)}_\alpha$ are chosen such that they minimize the
effective mass
\be
m_{eff}(t_C)=-{1 \over t_C}\log\left[
{v^{(R)}_\alpha v^{(R)}_\beta {\mathbf C}_{\alpha\beta}(t_C)
\over
v^{(R)}_\alpha v^{(R)}_\beta {\mathbf C}_{\alpha\beta}(0) }
\right] \;\;,
\ee
where $t_C$ is time  separation for the optimization and repeated
indices are summed over. In our simulation
$t_C=1$ is taken. If we denote the optimal values of $v^{(R)}_{\alpha}$
by a column vector ${\mathbf v}^{(R)}$,
this minimization is equivalent to the following eigenvalue problem:
\be
{\mathbf C}(t_C)\cdot {\mathbf v}^{(R)}
=e^{-t_Cm_{eff}(t_C)}{\mathbf C}(0)\cdot {\mathbf v}^{(R)}
\;\;.
\ee
The eigenvector ${\mathbf v}^{(R)}_0$ with the lowest effective
mass then yields the coefficients $v^{(R)}_{0\alpha}$ for
the operator ${\mathcal G}^{(R)}_0(t)$ which best overlaps the lowest
lying glueball in the channel with symmetry $R$.
Higher-mass eigenvectors of this equation will then overlap
predominantly with excited glueball states of a given
symmetry channel.

With these techniques, the glueball mass values are obtained in
\begin{table}[htb]
\caption{Glueball mass estimates for the symmetry channel
$A^{++}_1$, $E^{++}$ and $T^{++}_2$ at various lattice spacings.
The entries corresponding to the highest $\beta$ value are the
values after the infinite volume extrapolation. The last row
tabulated the continuum extrapolated result of the glueball mass
values in units of $1/r_0$.}
\vspace{3mm}
\begin{center}
\begin{tabular}{@{}llll}
\hline
$\beta$ &  $a_tM_{A^{++}_1}$ &  $a_tM_{E^{++}}$  &  $a_tM_{T^{++}_2}$ \\
\hline
 $2.4$ & $0.552(8)$ & $0.980(10)$  & $1.002(9)$ \\
 $2.6$ & $0.482(12)$ & $0.760(16)$  & $0.798(15)$ \\
 $3.0$ & $0.322(8)$ & $0.460(13)$  & $0.470(13)$ \\
 $3.2$ & $0.233(7)$ & $0.323(12)$  & $0.340(10)$ \\
\hline
 $\infty$ & $4.23(22)$ & $5.77(34)$  & $5.92(32)$ \\
\hline
\end{tabular}
\end{center}
\end{table}
lattice units and the final results are listed in Table.2. The
errors are obtained by binning the total data sets into several
blocks and doing jackknife on them.

\subsection{Extrapolation to the continuum limit}
\label{sec:extrap}
Finite volume errors are eliminated by
performing simulations at the same lattice spacing but different
physical volumes. This also helps to purge away the possible
toleron states whose energy are sensitive to the size of the volume.
A simulation at a larger volume is done for the smallest lattice
spacing in our calculation. We found that the mass of the scalar
glueball remains unchanged when the size of the volume is increased.
The mass of the tensor glueball  is  affected, which is
consistent with the known result that tensor glueballs have a
rather large size and therefore feel the finiteness of the
volume more heavily. The infinite volume is obtained by extrapolating
the finite volume results using the relation \cite{luscher86:finitea}:
\be
a_tM^{(R)}=a_tM^{(R)}_\infty
\left(1-{\lambda^{(R)} \over z}\exp(-{\sqrt{3}z \over 2})\right)\;\;,
\ee
where $z=M^{(A^{++}_1)}_\infty L_s$. Using the results for the mass of
the $E^{++}$ and $T_2$ glueballs on $8^324$ and $10^330$ lattices
for the same value of $\beta$, the final result for the mass of
these glueball states are obtained.
Glueball mass values for other symmetry sectors are not sensitive to
the finite volume effects. Therefore, in Table.2, only the extrapolated
values for the smallest physical volume are tabulated. Other entries
are obtained from $8^324$ lattice results.

As for the finite lattice spacing errors, special attention is paid
to the scalar glueball sector where the continuum limit extrapolation
was known to have problems. Due to the simulation points at small
lattice spacings, around $0.1fm$ and below, the ambiguity in this
extrapolation is reduced. We have tried to extrapolate the
result using different formula suggested in Ref.~\cite{colin99},
the extrapolated results are all consistent within statistical errors.
For definiteness, we take the simple form:
\be
\label{eq:extrap}
M_G(a_s) = M_G(0) 
+ {c_1 \over r_0}\left({a_s \over r_0}\right)^2 
+ {c_2 \over r_0}\left({a_s \over r_0}\right)^4\;\;,
\ee
and
\begin{figure}[htb]
\begin{center}
\includegraphics[width=7.5cm]{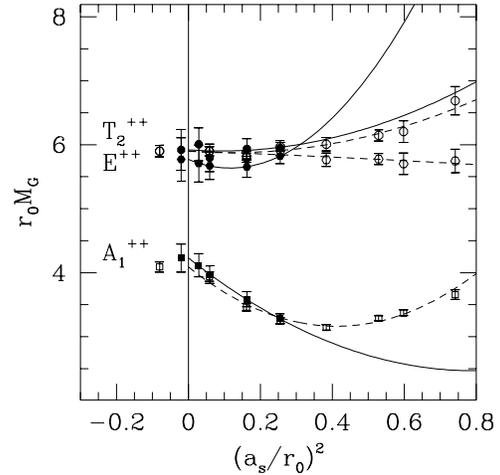}
\caption{The continuum limit extrapolation of glueball mass values
in scalar and tensor channels. The solid symbols are results from
this calculation with the corresponding continuum limit
extrapolation represented by the solid lines. For comparison, the
corresponding results from \cite{colin97,colin99} are also shown with
open symbols and dashed lines.}
\end{center}
\end{figure}
the result is illustrated in Fig.2. The final extrapolated results
for the glueball mass values are
also listed in Table.2.
The data points from our simulation results are shown
with solid symbols and the corresponding
extrapolations are plotted as solid lines.
It is also noticed that the extrapolated mass values for
$E^{++}$ and $T^{++}_2$ channels coincide within statistical
errors, indicating that in the continuum limit, they form
the tensor representation of the rotational group.
For comparison, corresponding results from \cite{colin97} are
also shown with open symbols and the dashed lines.
If one would  extrapolate linearly
in $(a_s/r_0)^2$ using three data points with
smallest lattice spacing, the results are statistically consistent
with the results using the extrapolation~(\ref{eq:extrap}) within
errors. 
Our final extrapolated result for the scalar glueball mass
lies higher, though still statistically
consistent, than that of Ref.7. This has to do with the fact that
simulation results at small lattice spacings are higher than those
at larger lattice spacings.
If one would extrapolate linearly using only the two data points with
the smallest lattice spacing from \cite{colin97,colin99},
one would arrive at a higher scalar glueball mass value
and closer to our result. This indicates that
the difference between our final result on the scalar glueball mass
and that of \cite{colin97,colin99} basically comes from
the ambiguity of the extrapolation when one tries to include
more data points from larger lattice spacings.
In our calculation, with the help of one more point at
smaller lattice spacing, using three data points with smallest
lattice spacings, one can extrapolate linearly towards the continuum
and obtain a result consistent with the result using
a quadratic extrapolation. In the tensor channel, a constant extrapolation
will also yield a consistent result.
It is seen that, due to data points at lattice spacings
around $0.1fm$ and below, the uncertainties in the extrapolation
for the glueball mass values are reduced.

Finally, to convert our simulation results on glueball
masses into physical units, we use the
result $r^{-1}_0=410MeV$. The errors for the hadronic
scale $r_0$ is neglected. For the scalar glueball
we obtain $M_G(0^{++})=1730(90)MeV$. For
the tensor glueball mass in the continuum, we
combine the results for the $T^{++}_2$ and
$E^{++}$ channels and obtain $M_G(2^{++})=2400(95)MeV$ for
the tensor glueball mass.

\section{DISCUSSIONS AND CONCLUSIONS}
\label{sec:conclusion}

We have studied the glueball spectrum at zero momentum in
the pure $SU(3)$ gauge theory using
Monte Carlo simulations on asymmetric lattices with the
lattice spacing in the spatial directions ranging from
$0.08fm$ to $0.25fm$. This helps to make extrapolations
to the continuum limit with more confidence
for the scalar and tensor glueball states.
The mass values of the glueballs are converted to physical
units in terms of the hadronic scale $r_0$.
We obtain the mass for the scalar glueball
and tensor glueball to be:
$M_G(0^{++})=1730(90)MeV$ and $M_G(2^{++})=2400(95)MeV$.
It is interesting to note that,
around these two mass values, experimental glueball candidates exist.
Of course, in order to compare with the experiments
other issues like the mixing effects and the 
effects of quenching have to be studied.

\end{document}